\documentclass[12pt]{iopart}

\usepackage{lineno}

\usepackage{amssymb}

\usepackage{txfonts}

\usepackage[pdftex,colorlinks=true,bookmarks=false,citecolor=blue,urlcolor=blue]{hyperref} 

\usepackage{graphicx}% Include figure files
\usepackage{dcolumn}% Align table columns on decimal point

\usepackage{multicol} %this errs if I want to set onecolumn. M.

\usepackage{CJKutf8}

\usepackage{bbm}

\begin{document}

%\linenumbers
\begin{CJK}{UTF8}{gbsn}
%\begin{CJK*}{GB}{gbsn}

\title{Angular momenta in fields from a rotational mechanical antenna}

\author{Yu Mao$^1$, Y. Liu (刘泱杰)$^{2, 3\#}$ and Hai Lin$^{1*}$}

\address{$^1$College of Physical Science and Technology, Central China Normal University, 152 Luoyu Road, 430079 Wuhan}
\address{$^2$School of Physics and Electronic Science, Hubei University, 368 Youyi Avenue, 430062 Wuhan}
\address{$^3$Lab of Micro-Nano Devices and Microsystems, School of Electronics and Information, Hangzhou Dianzi University, 310018 Hangzhou}

\ead{\mailto{Correspondence authors  $^{\#}$yangjie@hubu.edu.cn, $^{*}$linhai@mail.ccnu.edu.cn}. }

\begin{abstract}
\noindent Mechanic antennas provide opportunities for human portable, VLF communications, where a rotational dipole emits EM signals with angular momenta. In this paper we analytically derive the electromagnetic fields from a rotational electric dipole using Fourier transform method, and find that the radiated fields from the rotational electric dipole carries nonzero energy flow density in both orbital and spin angular momentum (AM) parts by AM flux tensors. Intuitively, a rotation of a dipole induces a longitudinal orbital angular momentum and a longitudinal spin angular momentum both circulating in the rotation direction. And the binding force for the rotational electric dipole is then shown to result mainly from the Coulomb fields. We believe that our work can provide novel communication designs for portable mechanic antennas. 

%Keywords: Geometrical optics, Optical system design, Imaging and optical processing, Edge and boundary effects, Gradient-index lens

%\pacs{42.15.-i, 42.15.Eq, 42.30.-d, 42.25.Gy}

\end{abstract}

\footnote {Submitted to \JPD on 13th May, transferred to \emph{J. Phys. Commun.} 3rd August, 2021.}

% re-revised 14 Aug-2 Sept, \\
%re-submitted 2 Sept 2015; appeared in ArXiv 1505.01534v2. }
%{Rejected by {\itshape Phys. Plasmas}(05-Oct-2013),\\
 %{\JPD} \\Received (11-Oct), moderate revision suggested(08-Jan-2014), re-submitted 16-Jan (Article reference: JPhysD-100502.R1), 
 %Accepted for publication 27-Jan, proofread 15-Feb, 
 %Published online 6-March-2014. \\
 %Please cite this Paper as: J. Phys. D: Appl. Phys. 47 (2014) 125502. \url{http://iopscience.iop.org/0022-3727/47/12/125502/}}

\maketitle

\end{CJK}

\section{Introduction}

Traditional electromagnetic radiators are formed by using voltage to drive current in conductive antenna elements. The physical size of the communication system are highly relevant to the antenna size which are bounded by the Wheeler-Chu limit. This restriction hampers the usage of radio frequency technologies in specific communication requirements. For instance, at the very low frequency (VLF) band, whose wavelengths are 10-100km long, even an electrically small antenna will require a relatively large structure size. When constrained in size, traditional VLF antennae are very inefficient.

Mechanic antennas, differing from convectional antennas, promise opportunities of human portable platforms for VLF communications. By means of physically moving or oscillating an electric dipole or magnetic dipole, radiated electromagnetic signals can be generated from different kinds of mechanical motion. Due to its low-loss merit, the mechanical antenna system has the possibility to break Chu's limit which can be used as a compact transmitter in the VLF range. Such mechanical transmitters can be made from a rotational electret or magnet~\cite{Bickford2019}.

On the other hand, if we simply imagine a rotational dipole, i.e. a spinning charge pair, it will emit out electromagnetic (EM) waves continuously~\cite{Jackson1999}. The rotation of this kind of wave emitter (dipole) on the other hand, is reminiscent of a vortex geometry intuitively, considering the helical phase of the orbital angular momentum for an electromagnetic wave beam~\cite{Allen1992, Bliokh2014b, Bliokh2015b, Bliokh2015c, Barnett2016}. This extra degree of freedom on orbital angular momenta may well provide additional channels to modulate signals which would greatly promote communication capacity, knowing that in principle no upper limit of topological orders for orbital angular momenta exist. This inspires our question: whether rotation of charge can possibly generate any type of EM vortex beam, i.e. with either orbital or spin angular momenta? In this paper, we derive analytically the EM fields for a slow rotational dipole, and find that orbital angular momentum (OAM) is generated only in the longitudinal direction and that the spin one (SAM) also appears in the longitudinal direction. The rotational rate is assumed to be non-relativistic in consideration of realistic mechanical rotators, which differ from relativistically-moving charge to generate synchrotron radiation~\cite{Hofmann2007, Hacyan2016}. We also find that the binding force to counteract the mutual attraction mainly results from the Coulomb field.

Our model considers electromagnetic fields from a rotating electric (magnetic) dipole, which is the prototype of a rotational dipole antenna to radiate. In this paper, we explore the orbital and spin angular momenta, and the reaction force for a rotational dipole. This type of rotational dipole model may apply to a very low frequency ($3\sim30$kHz) portable electrically-small antenna~\cite{Pfeiffer2017, Kemp2019, DongC2020} making use of a mechanical rotation to generate EM waves with OAM and SAM, instead of relying on the interaction of incident waves with solely artificial micro-structures~\cite{DuL2017, ShiP2018, ChenM2018, ShiH2019, Mock2018, Hassanien2019}. Considering the potential application to encode the angular momenta of EM waves in wireless communication, our simple model could provide some helpful insights for its engineering design.

The main structure of the rest of our paper is as follows. In Sec.~\ref{II}, we use the multipole expansion method to describe our rotational dipole and give its explicit fields in form of periodic impulses in frequency domain. In Sec.~\ref{AM}, the OAM and SAM for the dipole fields are analyzed from the perspectives of Poynting vector and electric fields. Sec.~\ref{reaction} calculates the binding force for rotational dipole. Then we summarise the paper in Sec.~\ref{sum}.

\section{Fields calculated from the multipole description}\label{II}

\begin{figure*}[htbp!]

%%\includegraphics[width=0.37\textwidth]{Electret.png}\\
%\begin{tikzpicture}
 % %\draw[step=.5cm,gray,very thin] (-1.4,-1.4) grid (1.4,1.4);

%\node at (-2.5,2.6)  {(a)} ; 
  %\draw [->, thick](-2.3,0) -- (2.3,0); \node [right=68pt] {$x$} ; 
  %\draw [->, thick](0,-2.3) -- (0,2.3);  \node [above=68pt] {$y$} ; 
%\node [left=70pt] {$d$} ; 
 % \draw (0,0) circle (2.cm); 
  %\draw [->](20mm,14mm) arc (0:30:6mm);\node [above right=38pt] {$\Omega$};
  %\node at (0,2) [circle,draw=black!10, fill=blue!30] {$+$};
  %\node at (0,-2) [circle,draw=black!10,fill=red!30] {$-$};
    %\draw (-2.2,2) -- (-1.8, 2); 
       % \draw[yellow, very thick] (-0,-1.73) -- (-0, 1.73); 
        %\draw (-2.2,-2) -- (-1.8, -2); 
         % \draw [<->, ](-2.,2.) -- (-2.,-2.); 
%\end{tikzpicture}\\
%\begin{tikzpicture}
%\node at (-2.5,2.6)  {(b)} ;
 % \draw [->, thick](-2.3,0) -- (2.3,0); \node [right=68pt] {$x$} ; 
  %\draw [->, thick](0,-2.3) -- (0,2.3);  \node [above=68pt] {$y$} ; 
%\node [left=70pt] {$d$} ; 
 % \draw (0,0) circle (2.cm); 
  %\draw [->](20mm,14mm) arc (0:30:6mm);\node [above right=38pt] {$\Omega$};
 %\shade (-0.3,2) rectangle (0.3,-2) ; %(0.5,-2) ;%circle (.5cm);
   %\node at (0,1.7) [,draw=black!10, fill=blue!30] {N};
  %\node at (0,-1.7) [,draw=black!10, fill=red!30] {S};
    %\draw (-2.2,2) -- (-1.8, 2); 
      %  \draw (-2.2,-2) -- (-1.8, -2.); 
        %  \draw [<->, ](-2.,2) -- (-2.,-2.); 
%\end{tikzpicture}

\includegraphics[width=0.9\textwidth]{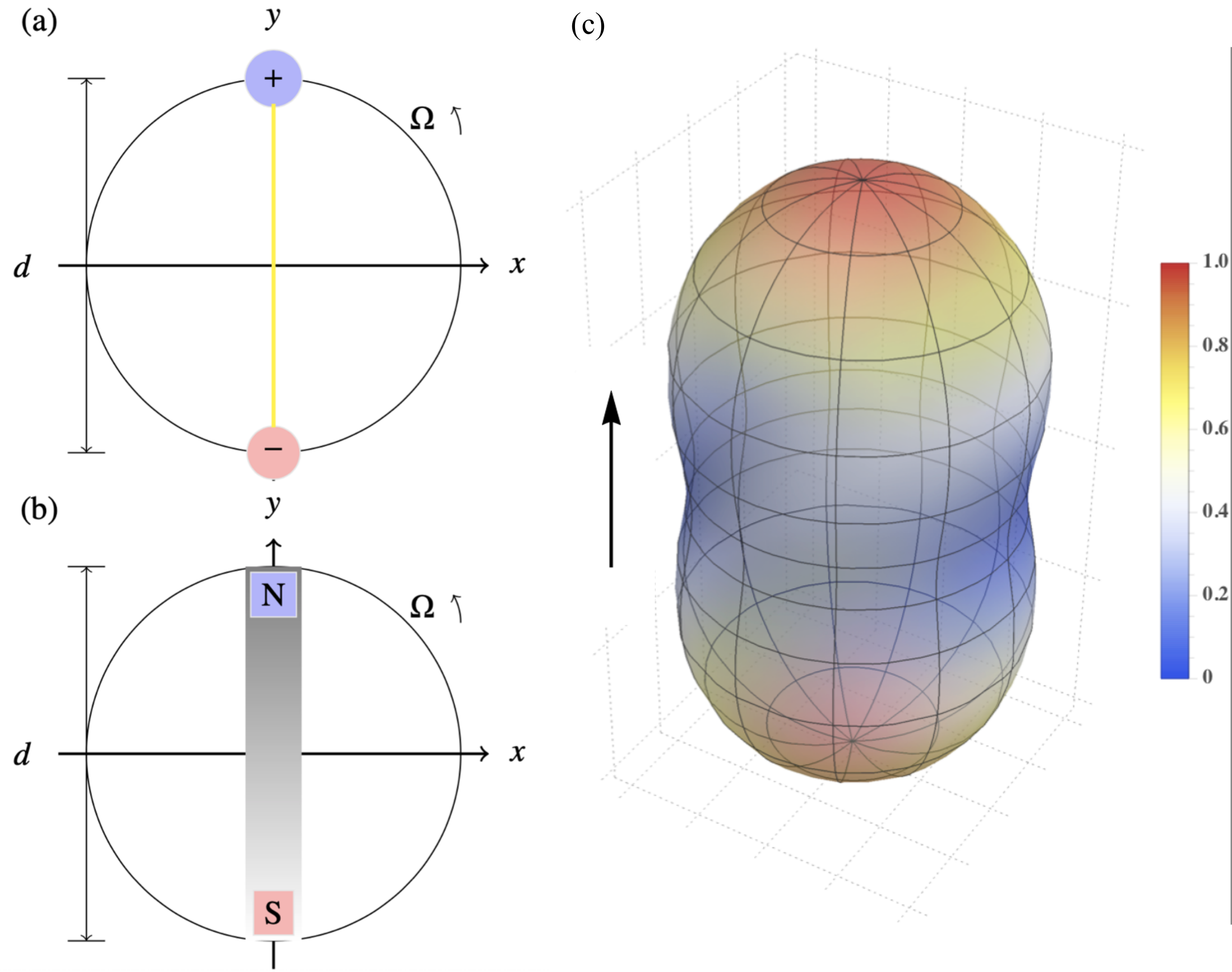}
\caption {\label{fig1} (a-b) Schematic for (a) rotational electric dipole and (b) rotational magnetic dipole;  (c) Its far field radiation pattern with the black arrow for z direction. (a-b) produced from \LaTeX~\texttt{tikz} and (c) produced from \texttt{Mathematica}~\cite{Mathematica}. 
 }

\end{figure*}

\emph{Electric dipole}: For a dipole of length $d$ and charges $\pm q$ uniformly rotating at angular frequency $\Omega$ along z direction as shown in Fig.~\ref{fig1} (a), we choose the multipole-expansion method to derive the induced fields~\cite{Jackson1999} (Chap.~9, p407) to treat our problem, whose convenience will be shown later. We reasonably assume that for a rotational dipole, its charge and current $\tilde\rho(t), \tilde{\mathbf{J}}(t)$ can be Fourier-decomposed into $\rho(\omega), \mathbf{J}(\omega)$ in the frequency domain~\cite{Oppenheim2015}~(see Sec.~1 of supplement 1 for detail). 
%\textcolor{urlblue}{supplement 1} for detail). 

We will see quickly that the emitted field actually oscillates only at $\omega=\pm\Omega$. For an electric dipole rotating at frequency $\Omega$ (other forms of rotation such as radially-vibrating or other asymmetric ones similar to~\cite{Sarma2015, Ge2015} shall be further discussed elsewhere), the charge distribution can be written formally with aid of impulse functions in cylindrical coordinates $(r, \theta, z)$,
\begin{eqnarray}\label{rhot}
\tilde\rho(t)&=&q\frac{\delta(r-d/2)}{r}\Big[\delta\Big(\theta-\theta_+\Big)-\delta\Big(\theta-\theta_-\Big) \Big]\delta(z),\\
\Big(\theta_+:&=&\Omega t+\frac{\pi}{2},\theta_-:=\Omega t-\frac{\pi}{2}, r:=\sqrt{x^2+y^2},  \theta=\arctan2 [y, x]\Big). 
\end{eqnarray}
Here the impulse function $\delta(\cdot)$ is defined in the sense of $\int_{-\infty}^{\infty}{\rm d}x\delta(x-x_0)=1$~\cite{Bracewell2000}. Noting that in Eq.~\eref{rhot} the charge distribution is periodic in time domain with the minimum period $2\pi/\Omega$, as suggested in the arguments of impulse functions. Thus the charge distribution in Fourier domain can be written as
\begin{equation}\label{rhox}
\rho(\omega; r, \theta, z)=-\frac{iq}{\pi\Omega r}\delta(z)\delta\Big(r-\frac{d}{2}\Big)e^{i\frac{\omega}{\Omega}\theta}\sin\Big({\frac{\pi}{2}\frac{\omega}{\Omega}}\Big)\sum_{n=-\infty}^{\infty}\delta\Big(\frac{\omega}{\Omega}-n\Big), 
\end{equation}
which is derived in detail in Sec.~2 of {supplement 1} and contains the spatial-temporal term $\exp{i\omega\theta\Omega}$. The higher-order harmonic terms in the charge density are seen to vanish quickly in the dipole function. We then compute the fields from this charge distribution using the multipole expansion method.

For a rotational electric dipole, its electric dipole moment $\mathbf{p}$ explicitly vibrates monochromatically at the single rotation frequency $\Omega$ only,  
\begin{eqnarray}
\label{pomega}
\mathbf{p}=\frac{qd}{2}\Big[(-i\hat{x}+\hat{y})\delta(\omega-\Omega)+(i\hat{x}+\hat{y})\delta(\omega+\Omega)\Big],
\end{eqnarray}
after substituting Eq.~\eref{rhox} into the definition of electric dipole moment. The calculated electric dipole moment in Eq.~\eref{pomega} is based on understanding for charge distribution $\mathbf{x}_\mathbf{2}=(\hat{x}\cos\theta+\hat{y}\sin\theta)d/2$ along with Eq.~\eref{rhox}.  Here the spinning opposite charges contribute to the dipole moment equally, as only the positive spinning charge will occupy half of Eq.~\eref{pomega} after a similar derivation above ($\mathbf{p}_+=\mathbf{p}_-={\mathbf{p}}/{2}$). And the current density for the rotational electric dipole is straightforwardly
\begin{equation}
\mathbf{J}(\omega)=\hat{\theta}\frac{d}{2}\Omega \rho(\omega),
\end{equation}
in agreement with the continuity equation for charge and current.

We then turn to calculate the magnetic dipole and other higher order moments which shall in principle to take care the whole fields from a system of local charge and current. The merit of multipole expansion lies in the fact that our rotated dipole does not constitute any other dipole components. The magnetic dipole moment is calculated to be zero: $\mathbf{m}=0$, due to the impulse train at $\omega=n \Omega\,(n\in \mathbb{Z})$. Furthermore, the quadrupole moment vector $\mathbf{Q}(\hat{n})$ also vanishes exactly for the rotational electric dipole $\mathbf{Q}(\hat{n})=0$, where the direction norm satisfies the relation $\hat{n}=\mathbf{x}/\vert\mathbf{x}\vert$.  
We now know that from the perspective of emitted fields, a rotational electric dipole \emph{is} indeed an electric dipole \emph{only} and the electromagnetic fields from the rotational electric dipole is completely representable by an electric dipole $\mathbf{p}$ in Eq.~\eref{pomega}. Therefore the exact EM fields for our rotated electric dipole are available~\cite{Jackson1999, Bickford2019}:
\begin{eqnarray}
\label{Hxfull}
\mathbf{H}_\mathbf{p}&=&\frac{ck^2}{4\pi}{e^{ikR}}\Big(\frac{1}{R}+\frac{i}{kR^2}\Big)\hat{n}\times \mathbf{p}, \\
\label{Exfull}
\mathbf{E}_\mathbf{p}&=&\frac{e^{ikR}}{4\pi\epsilon_0}\Big[\Big(\frac{k^2}{R}+\frac{ik}{R^2}-\frac{1}{R^3}\Big) \mathbf{p}+\Big(-\frac{k^2}{R}-\frac{3ik}{R^2}+\frac{3}{R^3}\Big)(\hat{n}\cdot \mathbf{p}) \hat{n}\Big],
\end{eqnarray}
where $R:=\vert\mathbf{x}\vert=\sqrt{x^2+y^2+z^2}$.

\emph{Magnetic dipole}: Likewise for a rotational magnetic dipole in Fig.~\ref{fig1}(b) with moment density $M_0$ and voltage $V_0$, its only multipole component is its magnetic dipole
\begin{eqnarray}
\mathbf{m}
&=&\frac{M_0V_0}{2}\Big[(-i\hat{x}+\hat{y})\delta(\omega-\Omega)+(i\hat{x}+\hat{y})\delta(\omega+\Omega)\Big]. 
\end{eqnarray}
And its electric dipole and quadrupole moments both vanish, $\mathbf{p}=\mathbf{Q}=0$. Therefore the electric and magnetic fields for a rotational magnetic dipole are then completely due to its magnetic dipole moment only, explicitly 
\begin{eqnarray}
\label{Emfull}
\mathbf{E}_{\mathbf{m}}&=&-\frac{Z_0}{4\pi}\frac{k^2e^{ikR}}{R}\big(1-\frac{1}{ikR}\big)\hat{n}\times \mathbf{m}, \\
\label{Hmfull}
\mathbf{H}_\mathbf{m}&=&\frac{e^{ikR}}{4\pi}\Big[\Big(\frac{k^2}{R}+\frac{ik}{R^2}-\frac{1}{R^3}\Big)\mathbf{m}+\Big(-\frac{k^2}{R}-\frac{3ik}{R^2}+\frac{3}{R^3}\Big)(\hat{n}\cdot\mathbf{m})\hat{n}\Big];
\end{eqnarray}
where the vacuum impedance is defined as $Z_0:=\sqrt{\mu_0/\epsilon_0}$. 
Note that to obtain the magnetic dipole fields Eqs.~\eref{Emfull} and \eref{Hmfull} from the electric dipole ones Eqs.~\eref{Hxfull} and ~\eref{Exfull}, one only need make the substitutions $\mathbf{p}\rightarrow \mathbf{m}/c$, $\mathbf{E}_\mathbf{p}\rightarrow Z_0\mathbf{H}_\mathbf{m}$ and $Z_0\mathbf{H}_\mathbf{p}\rightarrow -\mathbf{E}_\mathbf{m}$. Then the far-field radiation patterns for a rotational electric dipole and a magnetic one are exactly the same. In the rest of our paper, we shall focus on the rotational electric dipole only and drop the subscripts for electric dipole $\mathbf{p}$ to avoid abundance.

%Their temporal counterparts can be written as 
%\begin{eqnarray}
%\tilde{\mathbf{H}}(t)=\int{\rm d}\omega H(\omega)e^{-i\omega t}=\frac{\Omega^2qd}{4\pi cR}{\hat{n}}\times\vec{s}, \\
%\tilde{\mathbf{E}}(t)=\int{\rm d}\omega E(\omega)e^{-i\omega t}=\frac{Z_0\Omega^2qd}{4\pi cR}\hat{n}\times\vec{s}\times \hat{n},
%\end{eqnarray}
%where a shorthand vector on xy plane $\vec{s}:=-\hat{x}\sin\Omega(t-{r}/{c})+\hat{y}\cos\Omega(t-{r}/{c})$ is used. 

%\subsection{Radiation fields and near fields for rotational electric dipole}

The electric dipole fields Eqs.~\eref{Hxfull} and ~\eref{Exfull} can be separate into two parts: the far-field part and the near-field one (see Eqs.~(S17-20) in Sec.~1 of {supplement 1}). In near region the electric field $\mathbf{E}^{NF}(\omega)$, the $r^{-2}$ term is deliberately excluded as it extends over to radiation region~\cite{Jackson1999}. We shall write the monochromatic fields in both far- and near-field regions explicitly. The far fields $\mathbf{E}(\omega), \mathbf{H}(\omega)$ in the following simple form:
\begin{eqnarray}
\label{Eomega2}
\mathbf{H}(\omega)&=&\vec{h}_+(\omega)\delta_++\vec{h}_-(\omega)\delta_-,\\
\label{Homega2}
\mathbf{E}(\omega)&=&\vec{e}_+(\omega)\delta_++\vec{e}_-(\omega)\delta_-. 
\end{eqnarray}
The near fields are then explicitly
\begin{eqnarray}
\label{Hnomega}
\mathbf{H}^{NF}(\omega)&=&\vec{h}_+^{NF}(\omega)\delta_+ +\vec{h}_-^{NF}(\omega)\delta_-,\\\
\label{Enomega}
\mathbf{E}^{NF}(\omega)&=&\vec{e}_+^{NF}(\omega)\delta_+ +\vec{e}_-^{NF}(\omega)\delta_-, 
\end{eqnarray}
with all the coefficients for fields explicit in Eqs.~(S22-35), Sec.~1 of {supplement 1}, and the shorthand forms $\delta_+:=\delta(\omega-\Omega)$ and $\delta_-:=\delta(\omega+\Omega)$ adopted.

The energy flow of far fields or the time-averaged Poynting vector (indicated by a bracket) for the rotational electric dipole is obtained after substitution of Eqs.~\eref{Eomega2} and \eref{Homega2}, 
\begin{eqnarray}\label{P}
\mathbf{P}:
&=&\Re[\vec{e}_+(\Omega)\times\vec{h}_-(-\Omega)+\vec{e}_-(-\Omega)\times\vec{h}_+(\Omega)]\\
\label{Poynt}
&=&\frac{Z_0\Omega^4q^2d^2}{32\pi^2c^2R^2}\hat{n}(1+\cos^2\theta_{\hat{n}}), 
\end{eqnarray}
where the azimuthal angle $\theta_{\hat{n}}$ and the polar angle $\phi_{\hat{n}}$  for direction norm $\hat{n}$ are defined in a spherical coordinate so that $x=R\sin \theta_{\hat{n}}\cos\phi_{\hat{n}}, y=R\sin \theta_{\hat{n}}\sin\phi_{\hat{n}}, z=R\cos \theta_{\hat{n}}$. This temporal-dependence of radiated power differs from $\sin^2\theta_{\hat{n}}$ for that of the convectional electron dipole, which is due to the global rotation setup in our case, other than the locally-vibrating structure of the convectional one. The derivation details are given in Sec.~3 of {supplement 1}. There are two points about the Poynting vector Eq.~\eref{Poynt} to take note. First, the time-averaged power radiated per unit solid angle 
\begin{eqnarray}
\frac{{\rm d}P}{{\rm d}\sigma}=\frac{Z_0\Omega^4q^2d^2}{32\pi^2c^2}(1+\cos^2\theta_{\hat{n}})
\end{eqnarray}
is shown in Fig.~\ref{fig1}(c). Second the radiated power for a rotational electric dipole is $P=\int{\rm d}\sigma{{\rm d}P}/{{\rm d}\sigma}={Z_0\Omega^4q^2d^2}/{6\pi c^2}$, which is exactly twice as that of the \emph{static} electric dipole for the same parameter $qd$~\cite{Jackson1999}. Just to give a magnitude of order for estimation purpose, the radiated power of a rotational electric dipole with parameters $\Omega=40\pi \;{\rm rad/s}, qd=3.4\times 10^{-4}{\rm C\cdot m}$ is on the order of ${\rm 0.0064pW}$.

Similar to the far-field Poynting vector, the near-field Poynting vector 
\begin{equation}\label{Pn}
\mathbf{P}^{NF}=\frac{q^2d^2\Omega}{16\pi^2\epsilon_0R^6}(-\hat{x}y+\hat{y}x).
\end{equation}
It is curious to note that the radiation fields power in Eq.~\eref{Poynt} flows in radial direction $\hat{n}$ while the near field counterpart in Eq.~\eref{Pn} circulates on the rotation plane of the dipole.

\section{Optical angular momenta: democratic OAM and SAM}\label{AM}
Now we are ready to look at the angular momenta due to the electromagnetic fields from a rotational electric dipole. We will first separate the power fluxes for both far and near fields of the rotational electric dipole, into orbital and spin flow densities, then in Subsec.~\ref{flux} determine the quantities of OAM and SAM flux contained in the rotational dipole, each of which we find possesses a half of the total angular momentum fluxes. Afterwards, in Subsecs.~\ref{OAM} and ~\ref{SAM} we plot the phase, amplitude and polarization for far electric fields to demonstrate how one measures the angular momenta of the rotational dipole fields.

For a rotational electric dipole, its extrinsic total angular-momentum density (time-averaged) $\mathbf{s}$ in the far-field region
\begin{equation}\label{m}
\mathbf{s}=\mathbf{x}\times\mathbf{P}=\mathbf{x}\times\frac{Z_0\Omega^4q^2d^2}{32\pi^2c^2}\frac{\hat{n}}{R^2}(1+\cos^2\theta_{\hat{n}}), 
\end{equation}
is generally nonzero and up to the choice of coordinates, and becomes intrinsic, to be invariant between transformation of coordinates after integration in space~\cite{Bliokh2015c}
\begin{equation}\label{Vm}
\int dV\mathbf{s}=\int dV\mathbf{x}\times\mathbf{P}. 
\end{equation}
We separate the power flux density, or Poynting vector, into the orbital flow density and the spin flow density~\cite{Berry2009, Bekshaev2011}
\begin{equation}\label{Psum}
\mathbf{P}=\mathbf{P}_o+\mathbf{P}_s, 
\end{equation}
of which the orbital flow part can be calculated as
\begin{eqnarray}
\mathbf{P}_o&=&\frac{c^2\epsilon_0}{2}\Im\big[\mathbf{E}^*\cdot\frac{\nabla\mathbf{E}}{\omega}\big]\\
&=&c^2\epsilon_0\Im\big[\vec{e}_+(\Omega)\cdot\frac{1}{-\Omega}\nabla\vec{e}_-(-\Omega)+\vec{e}_-(-\Omega)\cdot\frac{1}{\Omega}\nabla\vec{e}_+(\Omega)\big]\\
\label{Po}
&=&\frac{Z_0\Omega^4q^2d^2}{32\pi^2c^2}\Big[\frac{\hat{n}}{R^2}(1+\cos^2\theta_{\hat{n}})+\frac{c}{\Omega R^4}(-\hat{x}y+\hat{y}x)\Big]\sim\frac{1}{R^2}.  
\end{eqnarray}
Eq.~\eref{Po} is obtained by writing out the inner product for their temporally-averaged counterpart, via the same route to the radiation field Poynting vector in Eq.~\eref{P} in Sec.~3 of {supplement 1}. And the far-field spin flow part from Eq.~\eref{Psum} is
\begin{equation}\label{Pso}
\mathbf{P}_s=\mathbf{P}-\mathbf{P}_o
=\frac{Z_0\Omega^3q^2d^2}{32\pi^2c}\Big[\frac{1}{R^4}(+\hat{x}y-\hat{y}x)\Big]\sim\frac{1}{R^3}.  
\end{equation}
Note that in our case the spin momentum part $\mathbf{P}_s={c^2\epsilon_0}\big[\nabla\times\Im(\mathbf{E}^*\times{\mathbf{E}}/{\omega})\big]/4$~\cite{Berry2009} appears inaccurate due to neglecting the near-field power($\sim R^{-3}$). Put in another way, separation of orbital and spin AM exactly stands only in far region, but not so in source region, which results in the inaccuracy~(see its proof in Sec.~4 of {supplement 1}). In fact the above separation of orbital flow density and spin flow density is not the only feasible one\cite{Berry2009, Bekshaev2011}, and we choose the above one based on solely the electric fields because the electric field is more accessible to realistic experiment measurements other than the magnetic field. Similarly, according to the near field Poynting vector $\mathbf{P}^{NF}$ in Eq.~\eref{Pn}, the near-field AM density should be
\begin{equation}
\mathbf{s}^{NF}=\mathbf{x}\times\mathbf{P}^{NF}=\mathbf{x}\times\frac{q^2d^2\Omega}{16\pi^2\epsilon_0R^6}[\hat{x}(-y)+\hat{y}x], 
\end{equation}
which fades away from the circle since the near-field power flux goes $\sim R^{-6}$. Thus there is no necessity to look at their orbital and spin flow densities separately. As the forthcoming Subsecs.~\ref{OAM} and ~\ref{SAM} show, we shall see that OAM and SAM exist stably in the far fields of the rotational electric dipole.

It is quite indirect to read off angular momenta from the respective parts of OAM and SAM in Poynting vector which appear in an extrinsic AM density in Eq.~\eref{m}. Rather, the physical AM concept is revealed in the integral form of AM density in Eq.~\eref{Vm}. From the perspective of power fluxes, only the leading term of orbital flow part $\mathbf{P}_o$ Eq.~\eref{Po} \emph{appears} to be capable to contribute to the angular momentum density $\mathbf{s}$ in Eq.~\eref{m} because only $\mathbf{P}_o\sim1/R^2$, but we shall see quickly in Subsec.~\ref{flux} that both $\mathbf{P}_o$ and $\mathbf{P}_s$ actually contribute to the integral AM quantity from the perspective of AM flux tensors. And the electric field phase and polarization status in Subsecs.~\ref{OAM} and ~\ref{SAM} also reveal clearly to confirm the above statement. 

Instead, one could integrate in space the AM density to quantity the AM quantities of the fields of a rotational electric dipole. For instance, in our coordinates where the charges of the dipole rotates along the circle $\mathbf{x}_\mathbf{2}=(\hat{x}\cos\theta+\hat{y}\sin\theta)d/2$, angular momentum should vanish $\sim\int dS[R\hat{n}\times \mathbf{P}]=0$~\cite{Crichton2000} on a sphere of radius $r$ since Poynting vector $\mathbf{P}$ lies along the direction norm $\hat{n}$. However, this statement is wrong as the far fields in Eqs.\eref{Eomega2} and \eref{Homega2} ignore the intermediate radial component of $E_r\sim R^{-2}$, which actually contributes to $z$ component of AM~\cite{Heitler1936}. Moreover, the AM flux tensor serves in a better and explicit way to illustrate the quantity of angular momenta from fields, which will be used to quantify the OAM and SAM of fields from a rotational electric dipole. 

\subsection{AM flux tensors}\label{flux}

The AM density $\mathbf{s}$ is defined locally in space and one can define an AM flux on surface to express its conversation law, similar to continuity equation for linear momentum. 
%\begin{equation}
%\frac{1}{c}\partial_t[\mathbf{\tilde{S}}(t)]_l+\partial_k\tilde{T}_{lk}=0,
%\end{equation}
%where the temporal power flux is defined as $\mathbf{\tilde{S}}(t):=\mathbf{\tilde{E}}(t)\times\mathbf{\tilde{H}}(t)$ and Maxwell stress tensor $\overleftrightarrow{T}$ in component form
%\begin{equation}
%\tilde{T}_{lk}=\frac{1}{2}\delta_{lk}(\varepsilon_0\vert\tilde{\mathbf{E}}\vert^2+\frac{1}{\mu_0}\vert\tilde{\mathbf{B}}\vert^2)-\varepsilon_0\tilde{E}_l\tilde{E}_k-\frac{1}{\mu_0}\tilde{B}_l\tilde{B}_k. 
%\end{equation}
Following Barnett's definition~\cite{Barnett2002}, the $l$-component of AM density is expressed in the form of AM flux in continuity equation: 
\begin{equation}
\frac{1}{c}\partial_t[\mathbf{\tilde{s}}(t)]_l+\partial_k\tilde{M}_{kl}=0,
\end{equation}
of which $\mathbf{\tilde{s}}:=\mathbf{x}\times[\mathbf{E}(t)\times\mathbf{H}(t)]$ indicates the temporal total AM density, as compared with the temporally-averaged $\vec{m}$ in Eq.~\eref{m}, and AM flux density $\mathbf{\tilde{M}}$
%$\overleftrightarrow{M}
is defined in component form
\begin{equation}
\tilde{M}_{li}=\epsilon_{ijk}x_j\Big[\frac{1}{2}\delta_{kl}(\varepsilon_0\vert\tilde{\mathbf{E}}\vert^2+\frac{1}{\mu_0}\vert\tilde{\mathbf{B}}\vert^2)-\varepsilon_0\tilde{E}_l\tilde{E}_k-\frac{1}{\mu_0}\tilde{B}_l\tilde{B}_k\Big]. 
\end{equation}
For any AM source, its AM flux has nine components in general. Considering the spherical wave fronts of a rotational electric dipole, we thus compute its all three components of temporally-averaged AM flux density through a far-field sphere surface,  
\begin{equation}\label{Mrl}
M_{rl}=-\frac{\varepsilon_0}{2}\Re[(\mathbf{x}\times \mathbf{E}_\mathbf{p}^*)_l(\mathbf{E}\cdot \hat{n})]. 
\end{equation}
The total AM flux contained in the rotational dipole field then is determined from 
\begin{eqnarray}
\mathcal{M}_{rx}&=&\int_0^{\pi}\sin\theta d\theta\int_0^{2\pi}d\phi r^2M_{rx}=%\frac{ik^3}{12\pi\epsilon_0}(p_2^*p_1-p_1^*p_2), 
0, \\
\mathcal{M}_{ry}&=&\int_0^{\pi}\sin\theta d\theta\int_0^{2\pi}d\phi r^2M_{ry}=%\frac{ik^3}{12\pi\epsilon_0}(p_2^*p_1-p_1^*p_2), 
0, \\
\label{Mrz}
\mathcal{M}_{rz}&=&\int_0^{\pi}\sin\theta d\theta\int_0^{2\pi}d\phi r^2M_{rz}=%\frac{ik^3}{12\pi\epsilon_0}(p_2^*p_1-p_1^*p_2), 
\frac{k^2q^2d^2\Omega}{12\pi c\varepsilon_0}\delta_+\delta_-, 
\end{eqnarray}
and $\mathcal{M}_{rx}$, $\mathcal{M}_{ry}$ both vanish~\cite{Barnett2002}. So the angular momenta for the rotational dipole flow entirely in z-direction and none in other directions. 

To separate orbital AM and spin AM in Eq.~\eref{Mrz}, we adopt respectively two kinds of analytic definitions for OAM and SAM fluxes, which will turn out to indicate that OAM and SAM each possesses a half of total AM fluxes, consistent with Refs.~\cite{Crichton2000, Piccirillo2004}. First, using Yan's definition of SAM flux~\cite{YanS2019} and the exact electric fields Eq.~\eref{Exfull}, 
\begin{equation}
M_{s, ln}=\frac{\varepsilon_0}{2k^2}\Re[(\nabla\mathbf{E})\times\mathbf{E}^*]_{ln}=\frac{\varepsilon_0}{2k^2}\Re[(\partial_lE_k)\epsilon_{knp} E_p^*],
\end{equation}
we obtain three components of SAM flux on a far-field sphere as
\begin{eqnarray}\label{Msrz}
\mathcal{M}_{s, rx}&=&0,\\
\mathcal{M}_{s, ry}&=&0,\\
\mathcal{M}_{s, rz}&=&
\frac{k^2q^2d^2\Omega}{24\pi c\varepsilon_0}\delta_+\delta_-. 
\end{eqnarray}
Therefore the rest flux should be OAM flux possessing the other half of the total AM flux, and we have
\begin{equation}\label{Mrz2}
\mathcal{M}_{s, rz}=\mathcal{M}_{o, rz}=\frac{\mathcal{M}_{rz}}{2}. 
\end{equation}
Second, using Bliokh's definition of SAM and OAM flux~\cite{Cameron2012, Bliokh2014b} with exact fields Eqs.~\eref{Hxfull} and~\eref{Exfull}, we obtain the same results as above. We then conclude that the AM flux only propagates in longitudinal (z) direction which is split into two halves exactly as SAM and OAM flux. The significance of our using AM flux tensors lies in its explicit and exact forms for all three Cartesian components through spherical surface, which all results from the exact fields. This contrasts with Crichton's implicit and asympotic forms to arrive at the same conclusion as Eq.~\eref{Mrz2}, which is corroborated by our calculation.

We have two remarks to make in conclusion of Subsec.~\ref{flux}: (a) Firstly, our calculation of AM flux indicates that in far fields, though a sphere outside a rotational dipole the angular momenta of its fields flow completely in longitudinal i.e. $z$ direction. (b) Secondly, in Eq.~\eref{Mrl}, it is the radial component $\mathbf{E}\cdot\hat{n}$ that contribute to $z$ component of AM through the far-field sphere. (c) Only the far-field terms cannot constitute any AM flux for a rotational dipole, as the power flow separation in Eq.~\eref{Psum} considers only far fields and fails to elucidate the AM origins. It is the one-order-smaller term than far fields $-3ik(\hat{n}\cdot\mathbf{p})\hat{n}/R^2$ in Eq.~\eref{Exfull}, so-called the intermediate term, that plays a role in construction of AM for electromagnetic fields~\cite{Heitler1936}.

\subsection{Electric field plot: OAM for rotational electric dipole}\label{OAM}

\begin{figure*}[htbp!]
\begin{center}
\includegraphics[width=0.45\textwidth]{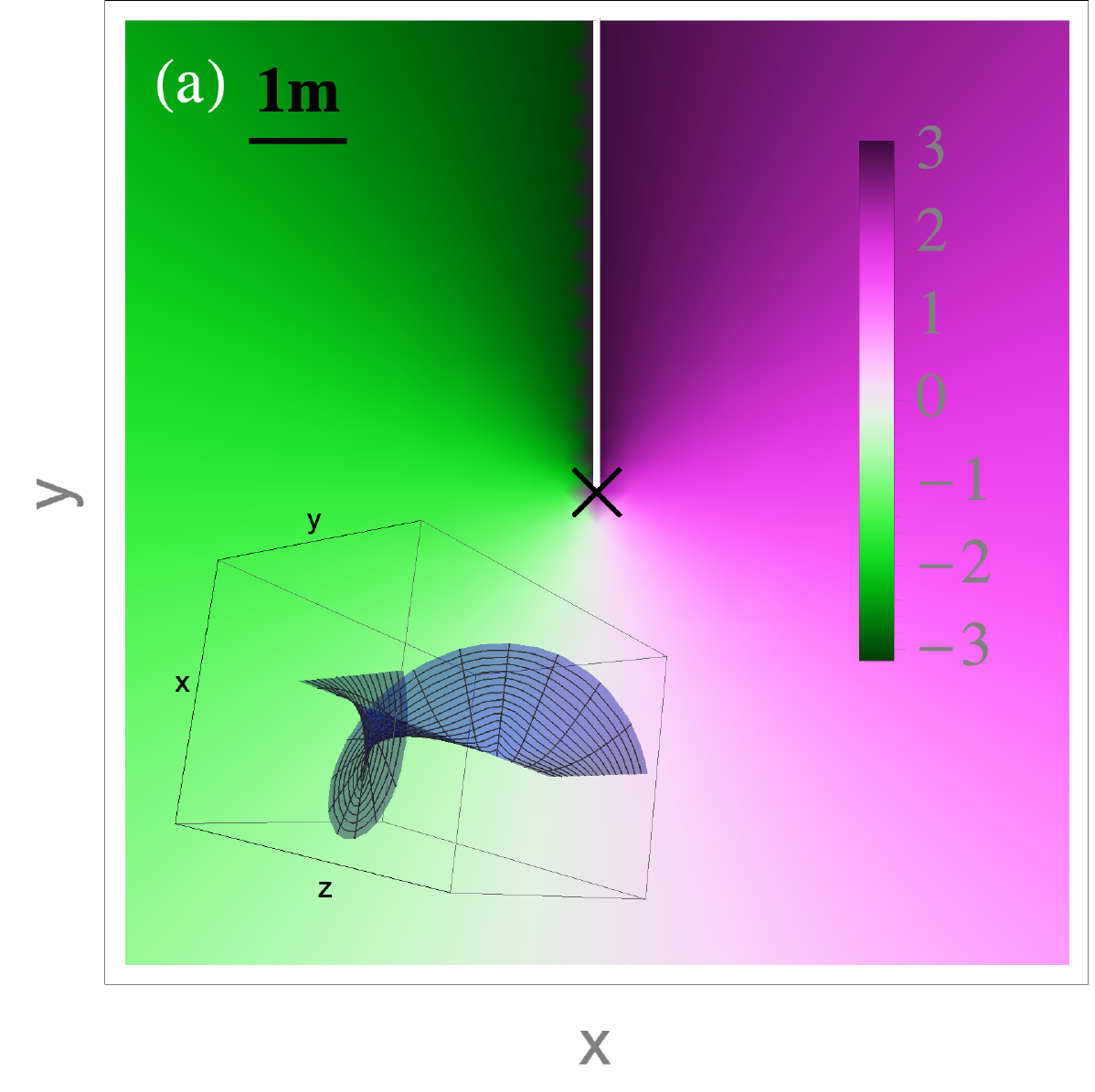}
\includegraphics[width=0.45\textwidth]{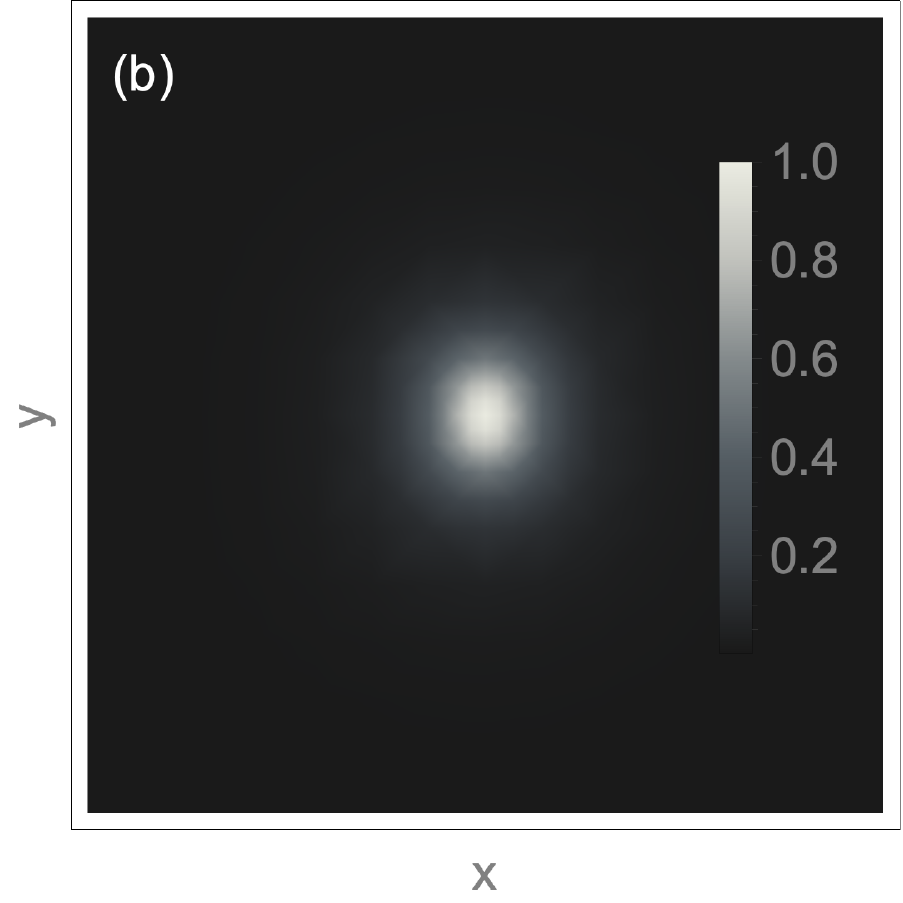}

\caption { \label{newfig3} (a) Phase plots for z components of far-field electric fields $E_z$ on plane $z=z_0$ respectively. Note the phase singularity point in panels (a) is the origin $(0, 0)$ (denoted by cross markers) and the branch cuts are denoted by white curves where the phase actually continues. Inset of (a) shows the helical phase of order 1 for far-region electric field. (b) The relative far-field amplitude distribution $\vert \mathbf{E}(x, y)\vert_{z=z_0}$. The scalebar in panel (a) is of a representative length $z_0=1\rm m$, which is actually scalable to give the same plots for other values of $z_0$. This also applies to also to panel (b) and Fig.~\ref{newfig5}. 
 }

\end{center}
\end{figure*}

For the orbital angular momentum in far field region, the amplitude and phase features of electric fields therein are plotted in Fig.~\ref{newfig3}. Only the phase of electric component $E_z$ is plotted in panel (a) because the transverse components of electric fields $E_x, E_y$ carry no singularity at all, of which the dominant term in Eq.~\eref{Exfull} is $-k^2(\hat{n}\cdot \mathbf{p})\hat{n}/R$. Panel (a) demonstrates that a phase singularity of $E_z$ with topological charge $l=1$ remains fixed at the origin even in the far field region, which confirms the existence of OAM for the far fields of rotational dipole.

\begin{figure*}[htbp!]
\begin{center}
\includegraphics[width=0.45\textwidth]{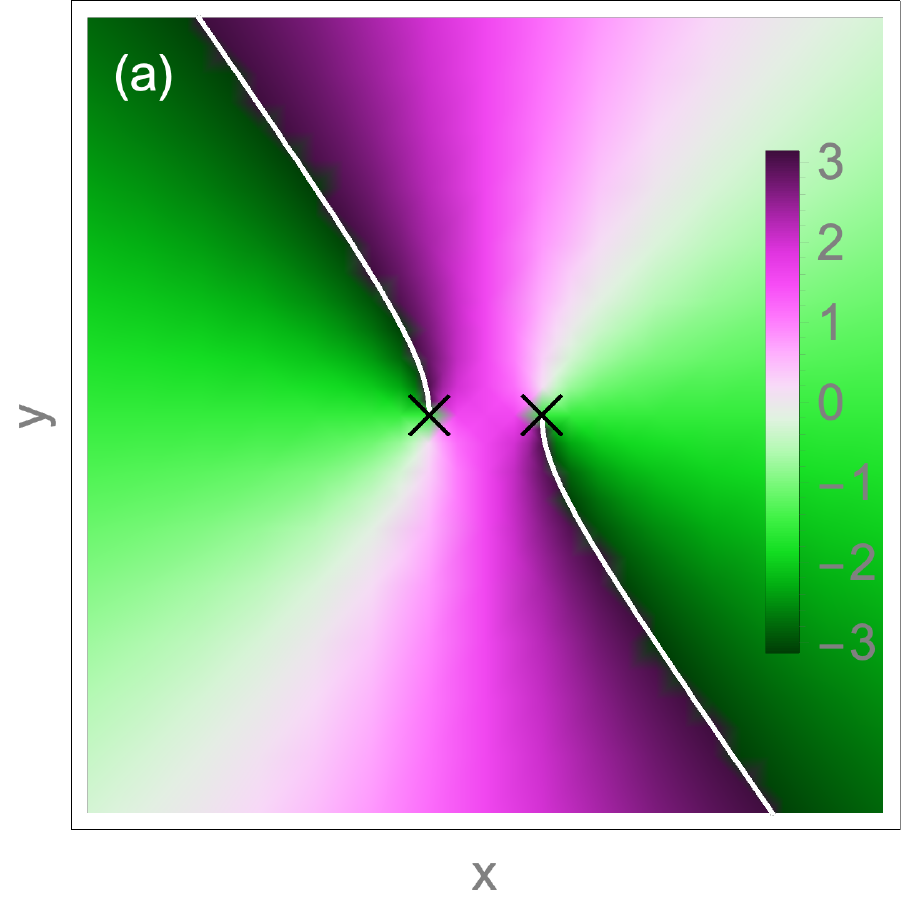}
\includegraphics[width=0.45\textwidth]{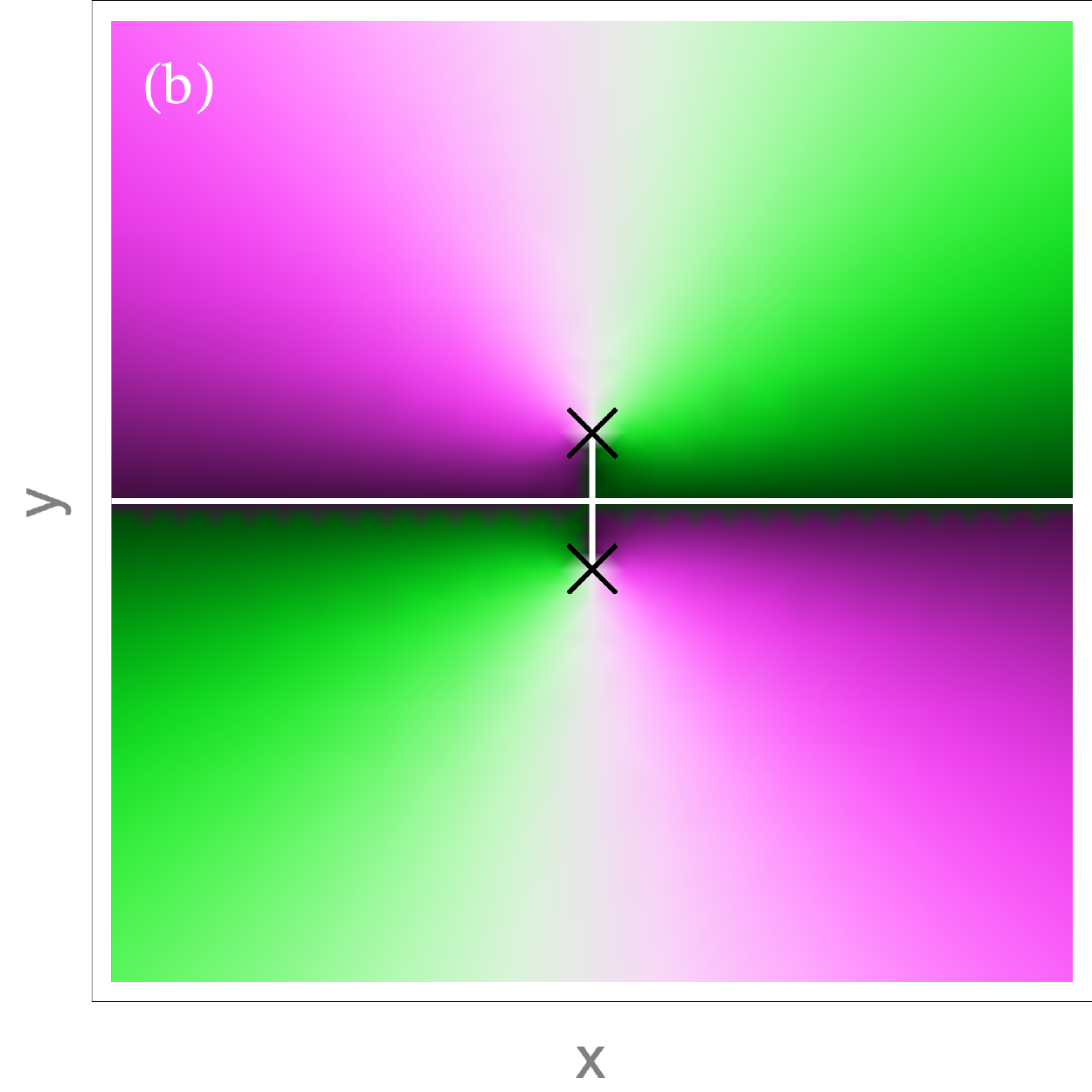}\\
\includegraphics[width=0.45\textwidth]{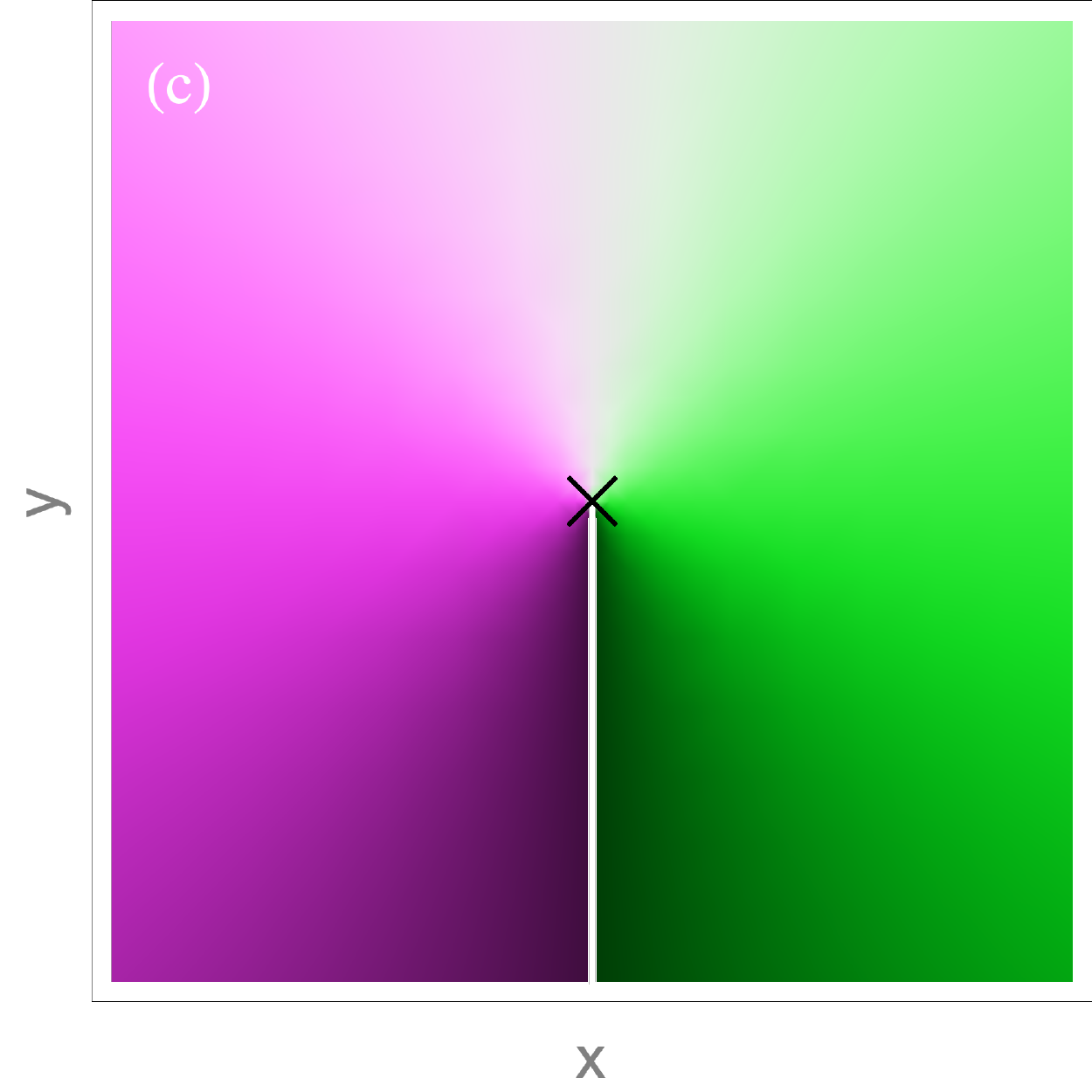}
\includegraphics[width=0.45\textwidth]{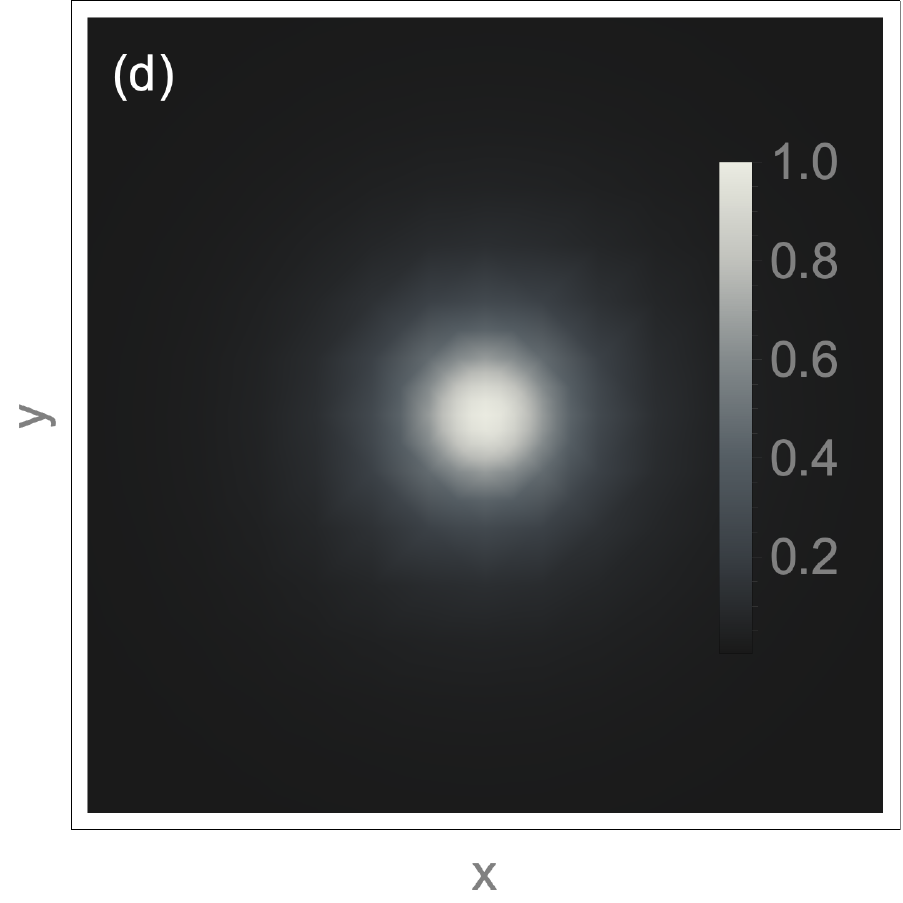}

\caption { \label{newfig5} (a-c) The phase plots for three Cartesian components of near-field electric fields: $E_x, E_y, E_z$ on plane $z=z_0$ respectively. Note the phase singularity points in panels (a-c) are respectively, $(\pm z_0/\sqrt{2}, 0),\; (0, \pm z_0/\sqrt{2}), $ and $(0, 0)$ (denoted by cross markers) and the branch cuts are denoted by white curves where the phase actually continues. The colorbars for (b-c) are the same as for (a). (d) The relative amplitude distribution $\vert \mathbf{E}(x, y)\vert_{z=z_0}$. 
 }

\end{center}
\end{figure*}

Moreover, it is also interesting to look at the orbital angular momentum in near field region, the amplitude and phase features of which therein are plotted in Fig.~\ref{newfig5}. The phase diagram differ between all three Cartesian electric components for near fields, as shown in panels (a-c) respectively. For the transverse near fields $E_x, E_y$, the orbital angular momenta carried by them are of topological charge $l=2$ while for $E_z$ of $l=1$, and the amplitude is radially decreasing as in panel (d). In panels (a-b), the phase singularities $(\pm z_0/\sqrt{2}, 0),\; (0, \pm z_0/\sqrt{2})$, respectively for $E_x, E_y$, will linearly move away from $z=0$ plane, and finally vanish in the far field region as $z_0$ increases to infinity, implying a conical singular region away from the origin. In panel (c), the phase singularity of $E_z$ instead remains fixed at the origin $(0, 0)$ as observers move away. Therefore only longitudinal component $E_z$ of the rotational electric dipole carries a phase singularity in the far field region. It is also noted that in general dipole radiation fields at large distance from the source is locally a plane wave, which does not carry OAM at all. The non-zero total OAM is a result of an integral all over the surface around the dipole. Therefore in our setup, the OAM detector should be located a certain distance $z=z_0$ away from origin ($z_0$ as an arbitrary distance) to measure the z-polarized electric field with a singularity.

The mixed order numbers of OAM in near fields described above indicates more complicated details of the dipole field phase than the simple helical phase known for Bessel beams~\cite{Allen1992}. Here to summarize Subsec.~\ref{OAM} we simply point out for a rotational dipole, only one phase singularity prevails in the longitudinal electric field, while those for the transverse components do not.

%\begin{figure*}[htbp!]
%\begin{center}
%\includegraphics[width=0.4\textwidth]{fig2a.png}
%\includegraphics[width=0.4\textwidth]{fig2b.png}

%\caption { \label{fig2} Vector plots for (a) far field orbital angular momentum part $\mathbf{P}_o$ and (b) near field counterpart at z=100m plane ($\Omega=40\pi\; {\rm rad/s}$). The black circle indicates the rotating trajectory of charge with the arrow indicating rotating direction. Produced from \texttt{Mathematica}. 
 %}

%\end{center}
%\end{figure*}

\subsection{Polarization states: SAM for rotational electric dipole}\label{SAM}

\begin{figure*}[htbp!]
\begin{center}
\includegraphics[width=0.5\textwidth]{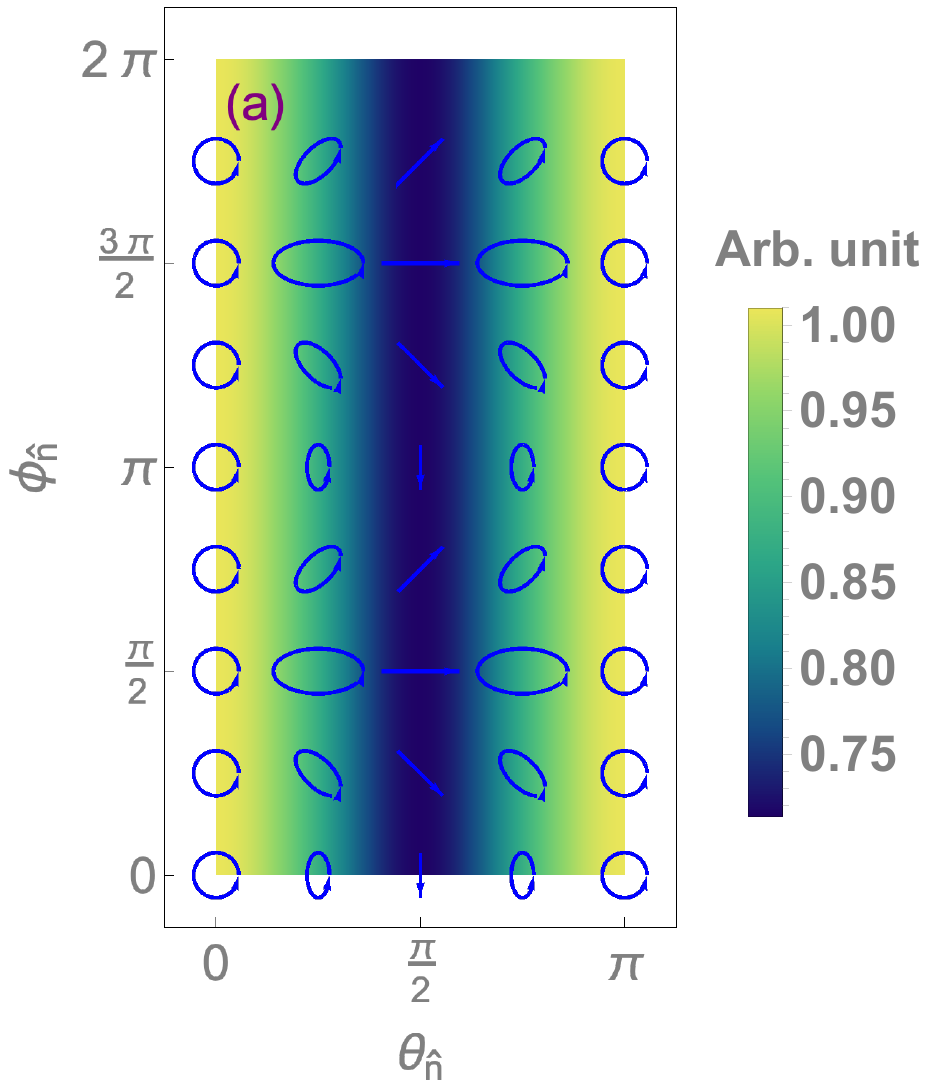}
\includegraphics[width=0.3\textwidth]{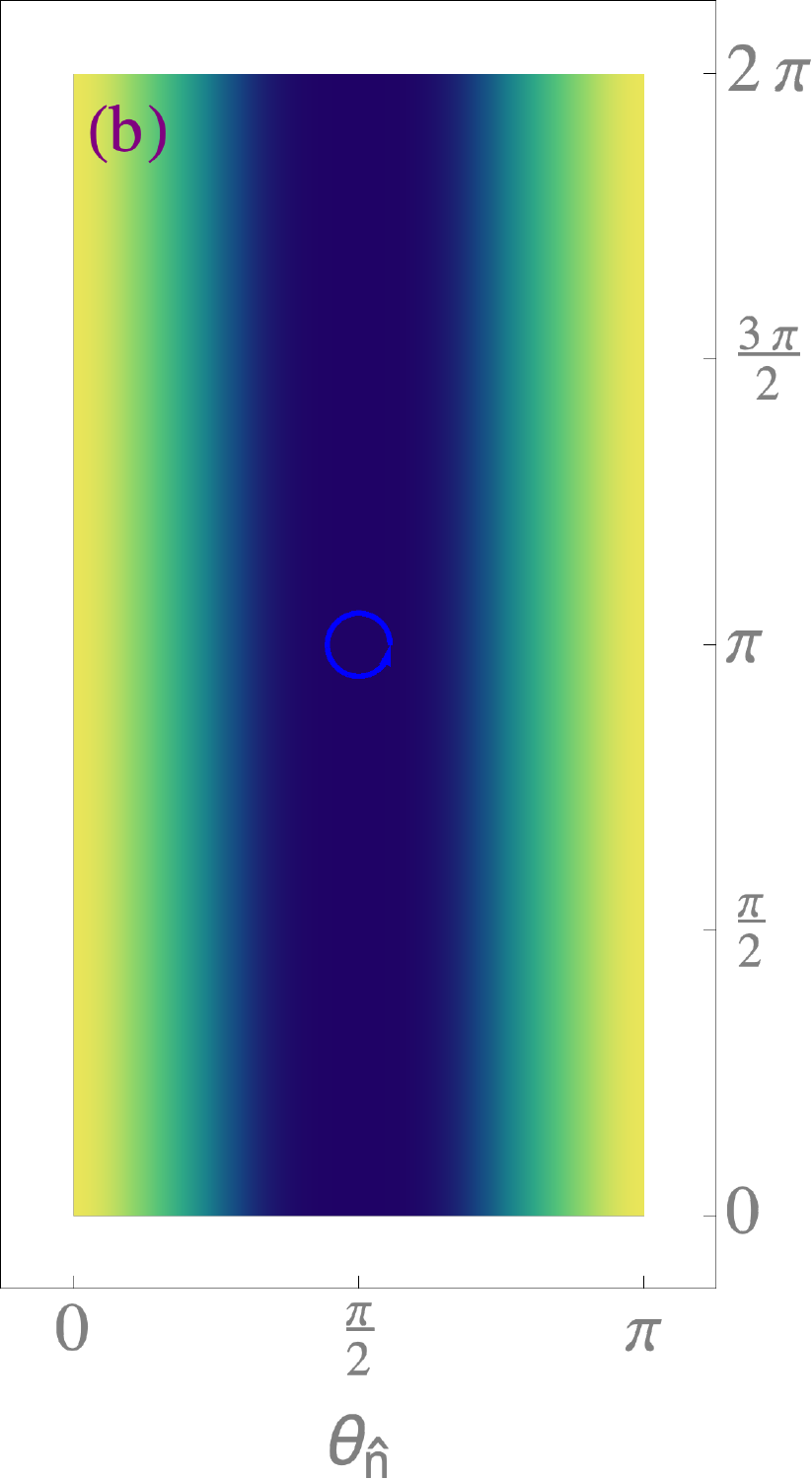}

\caption { \label{fig3} Relative magnitude distribution $\vert\mathbf{E}\vert(\theta_{\hat{n}},\phi_{\hat{n}})$ and $\vert\mathbf{H}\vert(\theta_{\hat{n}},\phi_{\hat{n}})$ superimposed by the corresponding polarization ellipses (in blue) for different direction angles of radiation fields: (a) electric fields  and (b) magnetic fields. (b) All polarizations for magnetic fields are uniformly left-circular and thus only one circle is plotted for clarity. 
 }

\end{center}
\end{figure*}

It is curious to look at the polarization details for the radiation fields of our rotated electric dipole. We use the classical Stokes parameters to compute the state of polarization from observation facing into the oncoming wave in z direction~\cite{BornWolf1999, Jackson1999}. Note that the polarization status varies generally differently and depends on the chosen observation direction. In this paper, only polarization status observed along $z$ axis is investigated. To be self-contained, we repeat the brief procedure to pinpoint the state of polarization for electric fields, as is experimentally more accessible compared to the magnetic fields. The Stokes parameters are~\footnote{Formally speaking, one should substitute instead $\mathbf{E}(\omega)$ in ~Eq.~\eref{Eomega2} to calculate the Stokes parameters. However, this route will lead to the same Stokes parameters as in Eqs.~\eref{s0}, \eref{s1}, \eref{s2} and \eref{s3}, except for an impulse factor $\delta_+^2\pm\delta_-^2$. We then take the simpler route as in Eqs.~\eref{s0}, \eref{s1}, \eref{s2} and ~\eref{s3} below. }
\begin{eqnarray}
\label{s0}
s_0&=&\vert\hat{x}\cdot \vec{e_+}(\omega)\vert^2+\vert\hat{y}\cdot \vec{e_+}(\omega)\vert^2,\\
\label{s1}
s_1&=&\vert\hat{x}\cdot \vec{e_+}(\omega)\vert^2-\vert\hat{y}\cdot \vec{e_+}(\omega)\vert^2,\\
\label{s2}
s_2&=&2\Re[(\hat{x}\cdot \vec{e_+}(\omega)) ^* (\hat{y}\cdot\vec{e_+}(\omega))],\\
\label{s3}
s_3&=&2\Im[(\hat{x}\cdot \vec{e_+}(\omega))^* (\hat{y}\cdot\vec{e_+}(\omega))],
\end{eqnarray}
in terms of the linear polarization basis $(\hat{x}, \hat{y})$. One then represents the electric field polarization as
\begin{eqnarray}\label{Exy}
E_x&=&\sqrt{\frac{s_0+s_1}{2}}\cos\Omega t,\nonumber\\
E_y&=&\sqrt{\frac{1}{2(s_0+s_1)}}\Big({s_2}\cos \Omega t+{s_3}\sin\Omega t\Big), 
\end{eqnarray}
based on which the polarization ellipses in Fig.~\ref{fig3} is then superimposed in the magnitude distribution of radiation fields for various direction angles $(\theta_{\hat{n}},\phi_{\hat{n}})$ in spherical coordinates. 

Our calculation shows that the polarization of electric and magnetic fields in all directions are left-hand polarized with different shapes of ellipses for observers facing the incoming wave along z direction. In Fig.~\ref{fig3}, (a) the polarization of electric fields are all left-hand polarized with various shapes of polarization ellipses and with a graded magnitude in $\theta_{\hat{n}}$ direction, while (b) that of magnetic fields are uniformly left-circularly polarized in all directions with also a similarly graded magnitude in $\theta_{\hat{n}}$. Specifically, on observation directions along lines $\theta_{\hat{n}}=0, \pi$, all electric fields are left circularly polarized and along line $\theta_{\hat{n}}= \pi/2$ are linearly polarized. This feature could be useful to measure the far-field hallmark of rotated dipoles in experiments. This polarization feature then corresponds to the spin AM part with the same sign as the orbital one in Eq.~\eref{Mrz2}, which can be pinpointed quantitatively by using degree of circular polarization(DoCP)~\cite{Olmos2019, Fernandez2012}.

We here note that both OAM and SAM for far fields of rotational dipoles are left-circular (cf.~Figs.~\ref{newfig3} and ~\ref{fig3}), which aligns with the left-circular rotation direction of the electric dipole itself (cf.~Fig.~\ref{fig1}(a) and Eq.~\eref{Mrz2}). For conclusion of Sec.~\ref{AM}, the rotation of a dipole induces an OAM and an SAM both in the longitudinal direction.

\section{Binding force of the rotational electric dipole}\label{reaction}

Knowing the rotational electric dipole emits both AM, it is then helpful to consider the binding force for such a setup, which shall counteract the electromagnetic interaction between the two point charges composing the dipole. In this Sec., we shall estimate the radiation reaction and also derive analytically the mutual interaction between the opposite charges related from the Li\'enard-Wiechart potential. We shall find that the EM interaction mainly results from the Coulomb attractive force between the charges.

\subsection{Reaction force}\label{reactionsub}
In this subsection, we calculate the reaction force of a rotational electric dipole and it turns out to be insignificant in our treatment. From the dynamic equation for a relativistic charged particle~\cite{Teitelboim2008, Jackson1999}, 
%\begin{eqnarray}
%\frac{{\rm d}P^\alpha}{{\rm d}\tau}=qF^{\alpha\beta}v_{\beta}-\frac{2}{3}q^2a^2v^\alpha. 
%\end{eqnarray}
one obtains its non-relativistic version in SI units
\begin{eqnarray}
m_0\frac{{\rm d}\mathbf{v}(t)}{{\rm d}t}+\frac{1}{6\pi\epsilon_0c^5}q^2a^2\mathbf{v}(t)=\mathbf{F}_{\rm ext}(t), 
\end{eqnarray}
which converts to 
\begin{eqnarray}\label{Momega}
\Big[-i\omega m_0+\frac{1}{6\pi\epsilon_0c^5}q^2a^2\Big]\mathbf{v}(\omega)=\mathbf{F}_{\rm ext}(\omega).
\end{eqnarray}
With the parameters $q/e=2\times 10^{17}\#, \Omega=40\pi  \,{\rm rad/s}, q=3.4\times 10^{-2}{\rm C}, d=1{\rm cm}$, one estimates the additional effective mass due to radiation reaction, as of 20 order of magnitude less significant than static mass ($\sim10^{-32} $vs. $\sim10^{-11} {\rm kg}$). Therefore radiation reaction force can be safely ignored in our problem of rotational dipole. For book-keeping we calculate the effective mass $M(\omega)$ following Eq.~\eref{Momega}, which implies to contribute only at the odd times the rotational frequency(see Sec.~5 of {supplement 1} for detail).

\subsection{Mutual electromagnetic interaction between the opposite charges} 
Apart from the radiation reaction, we need to calculate the electromagnetic force between the two rotating charges in our rotational dipole. One should use the retarded Li\'enard-Wiechart potential to compute the instantaneous forces instead of the multipole expansion one in Eqs.~\eref{Hxfull} and \eref{Exfull}. This is because the multipole expansion, essentially a Taylor expansion approximation route, fails to hold near the rotational curve $\vert\mathbf{x}_2\vert=d/2$.

Considering the circular trajectory $\mathbf{r}=-\hat{x}\frac{d}{2}\sin\Omega t+
\hat{y}\frac{d}{2}\cos\Omega t$ for the positive charge $+q$, which give its charge and current from Eq.~\eref{rhot} as  
\begin{eqnarray}
\tilde\rho_+(t)&=&q\frac{\delta(r-d/2)}{r}\delta\Big(\theta-\theta_+\Big)\delta(z), \\
\tilde{\mathbf{J}}_+(t)&=&\hat{\theta}\frac{d}{2}\Omega \tilde\rho_+(t);
\end{eqnarray}
we obtain the electric field exerted to charge $q$ by its partner $-q$ as 
\begin{eqnarray}
\mathbf{E}(\mathbf{r},t)=&\frac{-q}{4\pi\epsilon_0}\frac{\vert\mathbf{r}-\mathbf{r}_{\rm ret}\vert}{\big[c\vert\mathbf{r}-\mathbf{r}_{\rm ret}\vert-\frac{d^2}{4}\Omega\sin\Omega(t+t_r)\big]^3}\Big\{\Big[c^2+\frac{\Omega^2d^2}{4}\cos\Omega(t-t_r)\Big]\mathbf{u}-\\\nonumber
&\frac{\Omega^2 d}{2}[c\vert\mathbf{r}-\mathbf{r}_{\rm ret}\vert-\frac{d^2\Omega}{4}\sin\Omega(t+t_r)](-\hat{x}\sin\Omega t_r+\hat{y}\cos\Omega t_r)\Big\},
\end{eqnarray}
in which, 
\begin{eqnarray}
\mathbf{u}=&\hat{x}\Big[-\frac{cd}{2\vert\mathbf{r}-\mathbf{r}_{\rm ret}\vert}(\sin\Omega t+\sin\Omega t_r)-\frac{d}{2}\Omega \cos\Omega t_r\Big]+\\\nonumber
&\hat{y}\Big[\frac{cd}{2\vert\mathbf{r}-\mathbf{r}_{\rm ret}\vert}(\cos\Omega t+\cos\Omega t_r)-\frac{d}{2}\Omega \sin\Omega t_r\Big], \\
\mathbf{r}_{\rm ret}=&\frac{d}{2}(-\hat{x}\sin\Omega t_r+\hat{y}\cos\Omega t_r), 
\end{eqnarray}
and the retarded time $t_r$can be approximated by taking only the first leading term for the retarded time delay (cf.~Sec.~6 of ~{supplement 1})
\begin{equation}
t-t_r\approx \frac{d}{c}. 
\end{equation}
In our case of rotational dipole, the rotational velocity moves much slower than the light velocity $\Omega d/2\ll c$ and thus the leading terms with the higher order of $c$ for the electromagnetic fields become
\begin{eqnarray}
\mathbf{E}(t)&\approx& \frac{q}{4\pi\epsilon_0cd^2}\Big\{\hat{x}\Big[\frac{c}{2}(\sin\Omega t+\sin\Omega t_r)+\frac{\Omega d}{2}\cos\Omega t_r\Big]+\\\nonumber &&\hat{y}\Big[-\frac{c}{2}(\cos\Omega t+\cos\Omega t_r)+\frac{\Omega d}{2}\sin\Omega t_r\Big]\Big\}, \\
\mathbf{B}(t)&\approx&-\hat{z}\frac{\Omega q}{8\pi\epsilon_0c^2d}. 
\end{eqnarray}
Then the Lorentz force exerted to the positive charge should be 
\begin{eqnarray}
&&\int{\rm d}^3x[\tilde\rho_+\mathbf{E}(t)+\tilde{\mathbf{J}}_+\times\mathbf{B}(t)]\\
&\approx&\Big( \frac{q^2}{4\pi \epsilon_0 d^2}+\frac{q^2\Omega^2}{16\pi\epsilon_0c^2}\Big)(\hat{x}\sin\Omega t-\hat{y}\cos\Omega t),
\end{eqnarray}
where the first term from the electric field is less important from the second one from the magnetic field. So the Lorentz force is mainly 
\begin{eqnarray}
\int{\rm d}^3x[\tilde\rho_+\mathbf{E}(t)+\tilde{\mathbf{J}}_+\times\mathbf{B}(t)]\approx-\Big( \frac{q^2}{4\pi \epsilon_0 d^2}\Big)\hat{r}, 
\end{eqnarray}
where $\hat{r}=\hat{x}\cos\theta+\hat{y}\sin\theta=-\hat{x}\sin\Omega t+\hat{y}\cos\Omega t$ is used. We note that the electromagnetic field is dominated by the  attractive Coulomb force as if the charge were static. With the same parameters in Subsec.~\ref{reactionsub}, it is estimated that the Coulomb force is far larger than the acceleration force to maintain its rotation of the positive charge. Therefore the binding force for it should be \emph{almost} ${q^2}/{4\pi \epsilon_0 d^2}$, pointing \emph{almost} towards the rotation centre, in the same direction of the acceleration force $m\mathbf{a}(t)$ (order-of-magnitude estimation shows the Coulomb force overweighs the accerleration force). By the very same derivation, we also have 
\begin{eqnarray}
\int{\rm d}^3x[\tilde\rho_-\mathbf{E}(t)+\tilde{\mathbf{J}}_-\times\mathbf{B}(t)]\approx-\Big( \frac{q^2}{4\pi \epsilon_0 d^2}\Big)\hat{r}, 
\end{eqnarray}
for the negative rotating charge which goes as 
\begin{eqnarray}
\tilde\rho_-(t)&=&-q\frac{\delta(r-d/2)}{r}\delta\Big(\theta-\theta_-\Big)\delta(z), \\
\tilde{\mathbf{J}}_-(t)&=&\hat{\theta}\frac{d}{2}\Omega \tilde\rho_-(t). 
\end{eqnarray}

Therefore in this Sec., our calculation shows that for a rotational electric dipole, reaction force is insignificant and the binding force to maintain the described rotation simply counteracts the Coulomb force upon the two charges, almost the same as electrostatics indicates. 

\section{Summary}\label{sum}
In summary, our paper analytically derives the electromagnetic fields due to a rotational electric dipole using the multipole expansion method. We find that a rotation of a dipole induces fields entirely from an electric dipole, and that an orbital angular and a spin angular momentum both in the longitudinal direction, which aligns with straightforward intuition. And at least in a non-relativistic rotation scenario, the binding force to maintain the described rotation is calculated via the retarded potential to counteract \emph{almost} the Coulomb force on the two charges, the same as electrostatics indicates. We believe that our work can provide additional inspirations to generate angular momenta via simple mechanic setups.

\ack
We were supported by Fundamental Research Funds for the Central University of China (CCNU18JCXK02, 18GF006, 16A02016 and 19TS073); Open project of Guangxi Key Laboratory of Wireless Wideband Communication and Signal Processing (GXKL06190202); Electrostatic Research Foundation of Liu Shanghe Academicians and Experts workstation, Beijing Orient Institute of Measurement and Test (BOIMTLSHJD20181002); Hangzhou Dianzi University (ZX150204307002/023, KYZ043714070), Science and Technology Department of Hubei Province (2018CFB148); Natural National Science Foundation (NSFC11804087). We thank Ling Xiaohui, Chen Ke, Shi Hongyu, Du Luping, Wen Meng, Xu Yadong, and Referee 1's comments to improve our paper. 

See Supplement 1 for supporting content.

%\section*{Author contributions}
%H C proposed this starting idea and co-designed it with L Y. L Y performed \texttt{COMSOL} simulation, completed all the computation, and prepared the manuscript draft. H C revised the manuscript.  

%\begin{multicols}{1}  undefined control sequence. I.355 \col@number =1 PCLatex, M.

\newpage
\section*{References}

\bibliographystyle{unsrt}

\bibliography{bib1}

%\begin{thebibliography}{10}

%\href {\doibase 10.1088/1367-2630/11/9/093040} {\bibfield
  %{journal} {\bibinfo  {journal} {New J. Phys.}\ }\textbf {\bibinfo {volume}
  %{11}},\ \bibinfo {pages} {093040} (\bibinfo {year} {2009})}

%\bibitem{Yalunin2011}
%Yalunin S,  M and Ropers C 2011
%\newblock Strong-field photoemission from surfaces: Theoretical approaches.
%\newblock {\em Phys. Rev.} B {\bf 84}(19) 195426

%\bibitem{Liu2011}
%Liu Y and Ang L K May-22 2013
%\newblock Space Charge Effect of Time-dependent Emission Current Excited from Ultrafast Laser
%\newblock presented at {\em 14th IEEE International Vacuum Electronics Conference (IVEC)}, Paris, Poster Session {\bf I} 

%\end{thebibliography}

\noappendix

\newpage
%\Figures

%\end{multicols}

%%%%%%%now we start to write Supplemental Material. Y, 
%\pagebreak

%\onecolumngrid
\begin{center}
  \textbf{\large Angular momenta in fields from a rotational mechanical antenna: \\Supplement 1}\\[.2cm]
Y. Mao$^{1}$, Y. Liu$^{2, 3, \#}$ and H. Lin$^{1, *}$
 \\[.1cm]
  {\itshape ${}^1$College of Physical Science and Technology, Central China Normal University, 152 Luoyu Road, 430079 Wuhan\\
  ${}^2$School of Physics and Electronic Sciences, Hubei University, 430062 Wuhan\\
  ${}^3$Lab of Micro-Nano Devices and Microsystems, School of Electronics and Information, Hangzhou Dianzi University, 310018 Hangzhou}\\

  ${}^\#$Electronic address: yangjie@hubu.edu.cn\\
    ${}^*$Electronic address: linhai@mail.ccnu.edu.cn;

\end{center}
%\twocolumngrid

%\setcounter{equation}{0}
%\setcounter{figure}{0}
%\setcounter{table}{0}
%\setcounter{page}{1}
%\renewcommand{\theequation}{S\arabic{equation}}
%\renewcommand{\thefigure}{S\arabic{figure}}
%\renewcommand{\bibnumfmt}[1]{[S#1]}
%\renewcommand{\citenumfont}[1]{S#1}

This is a supplementary document for inclusion with submission to \emph{J. Phys. Commun.}. Dated: 13rd May, 2021.

\end{document}